# Observation of Electron-Hole Puddles in Graphene Using a Scanning Single Electron Transistor

J. Martin[1], N. Akerman[1], G. Ulbricht[2], T. Lohmann[2], J. H. Smet[2], K. von Klitzing[2], and A. Yacoby[1,3]

[1] *Department of Condensed Matter Physics, Weizmann Institute of Science, Rehovot 76100, Israel.*
[2] *Max-Planck-Institut für Festkörperforschung, Heisenbergstrasse 1, D-70569 Stuttgart, Germany.*
[3] *Department of Physics, Harvard University, Cambridge, MA 02138, USA.*

**The electronic density of states of graphene is equivalent to that of relativistic electrons [1-3]. In the absence of disorder or external doping the Fermi energy lies at the Dirac point where the density of states vanishes. Although transport measurements at high carrier densities indicate rather high mobilities [4-6], many questions pertaining to disorder [7-14] remain unanswered. In particular, it has been argued theoretically, that when the average carrier density is zero, the inescapable presence of disorder will lead to electron and hole puddles with equal probability. In this work, we use a scanning single electron transistor to image the carrier density landscape of graphene in the vicinity of the neutrality point. Our results clearly show the electron-hole puddles expected theoretically [13]. In addition, our measurement technique enables to determine locally the density of states in graphene. In contrast to previously studied massive two dimensional electron systems, the kinetic contribution to the density of states accounts quantitatively for the measured signal. Our results suggests that exchange and correlation effects are either weak or have canceling contributions**.

The kinetic energy of Dirac particles in graphene increases linearly with momentum [1,15,16]. The total energy per particle however, also referred to as the chemical potential, $\mu$, contains additional exchange and correlations contributions that arise from the Coulomb interaction $\mu = E_K + E_{ex} + E_c$ [17-22]. Here $E_K$, $E_{ex}$ and $E_c$ are the kinetic, exchange and correlation terms respectively. Therefore, the chemical potential and its derivative with respect to density, known as the inverse compressibility or density of states, provide direct insight into the properties of the Coulomb interaction in such a system. Compressibility measurements of conventional, massive, two dimensional electron systems made for example of Si or GaAs have been carried out by several groups [19,23-26]. It has been shown that at zero magnetic field the chemical potential may be described rather accurately within the Hartree-Fock approximation. Quantitatively it has been found that Coulomb interactions add a substantial contribution to the compressibility and become dominant at low carrier densities. In a perfectly uniform and clean graphene sample, the inverse compressibility is expected to diverge at the Dirac point in view of the vanishing density of states. This divergence, however, is expected to be rounded off by disorder on length scales smaller than our tip size. Long range disorder on the other hand will cause local shifts in the Dirac point indicative of a non-zero local density. In this work we measure the spatial dependence of the local compressibility versus carrier density across the sample. Fluctuations in the Dirac point across the sample are translated into carrier density maps through which quantitative information about the degree and length scale of disorder is obtained.

The preparation of graphene monolayers is performed in a similar manner as in Ref. [4,27]. We use Nitto tape to peel a large graphite flake from an HOPG crystal and press it onto a Si/SiO$_2$ wafer. Smaller and thinner pieces of this flake will then stick to the surface of the substrate. Once a suitable monolayer is selected in an optical microscope, two electrical contacts to the layer are patterned with optical lithography. Typical dimensions of our monolayers are $10 \times 4$ µm$^2$. The conducting Si substrate serves as a back-gate to vary the carrier density in the graphene sheet [4]. A back-gate voltage difference of 1V corresponds to a density change of $7 \cdot 10^{10}$ cm$^{-2}$. This conversion factor has been extracted from magnetotransport data (Fig. 1A). The unorthodox Hall quantization in graphene monolayers allows to distinguish them from multilayer flakes [5,6]. Fig. 1A shows the two terminal conductance $G_{2pt}$ as a function of density and magnetic field. Maxima in $G_{2pt}$ can be clearly seen at filling factors 2, 6, 10… for both electrons and holes. This behavior conclusively identifies the flake as a monolayer.

The local compressibility measurements of graphene described here are carried out using a scanable single electron transistor. Previous scanning tunneling microscopy (STM) studies on bulk graphite surfaces in strong magnetic fields already demonstrated the power of local methods to probe this class of materials [28-30]. However, unlike STM that probes the single particle density of states, the inverse compressibility measures the many-body density of states that includes information on the exchange and correlation energies as well. The experiments are performed at 0.3K. A schematic picture of the experimental setup is depicted in the inset to Fig. 1B. Details of our experimental method are described in Ref. [31,32]. The diameter of the SET is about 100 nm and the distance between SET and the sample is roughly 50 nm. The SET tip is capable of measuring the local electrostatic potential with microvolt sensitivity and a high spatial resolution close to its size. The inverse compressibility can be measured by monitoring the change in the local electrostatic potential, $\Phi_{total}$, when modulating the carrier density in the graphene flake with the back-gate, $\partial \Phi_{total}/\partial n$. We note that the small size of the graphene flake requires to consider position dependent direct pick-up from the fringing electric fields between the back-gate plane and the graphene flake as discussed in the methods section. In the absence of any transport current, any change in the local electrostatic potential is equal in magnitude and opposite in sign to the changes in the local chemical potential of the graphene, $e\frac{\partial \Phi_{total}}{\partial n} = -\frac{\partial \mu}{\partial n}$. Figure 1B shows the inverse compressibility for a fixed location of the SET as a function of the back-gate voltage at zero magnetic field. Based on the linear dispersion of the graphene bandstructure, the kinetic energy contribution to the inverse compressibility is expected to exhibit an unusual density dependence with a singularity at the neutrality point given by $\hbar v_F \cdot \sqrt{\frac{\pi}{|n|g_s}}$. Here $v_F$ is the Fermi velocity, $g_s$ is the band degeneracy, and $|n|$ is the carrier density measured from the neutrality point. At zero magnetic field $g_s = 4$ due to both spin and band symmetries. The maximum in the experimental $\partial\mu/\partial n$-trace clearly identifies the position of the Dirac point at this particular location. The absence of a singularity at the Dirac point in experiment is ascribed to disorder broadening. Since the kinetic term in graphene has the same density dependence as the leading exchange and correlations terms it is instructive to fit the data to the kinetic term with a single fit parameter that may be thought of as an effective Fermi velocity. The red line in Fig. 1B shows the result of such a fit and yields $v_F^{eff} = 1.1 \cdot 10^6 \pm 0.1 \cdot 10^6 \, m/s$. The indicated uncertainty reflects the accuracy with which the direct pick-up correction can be carried out. Since the Fermi velocity from band-structure calculations [1], $v_F = 1 \cdot 10^6 \, m/s$, lies within our experimental uncertainty, we conclude that the data can be described quantitatively by the kinetic term alone. It indicates that the exchange and correlation contributions are either weak or cancel out. Also theory has suggested only weak modifications of the compressibility due to exchange [20-22]. This is in marked contrast to massive two dimensional systems such as GaAs where at low densities the exchange term dominates [19,23-25]. Moreover, in these conventional two-dimensional electron systems quantitative agreement between experiment and theory is only accomplished when taking into account the non-zero thickness of the 2D layer [19,24]. The finite width is responsible for a softening of the Coulomb interaction between the electrons as well as a Stark like shift of the confinement energies in the potential well when the density is tuned. Both effects produce non-generic changes to the compressibility terms, which depend on the details of the heterostructure. These complications however do not arise for graphene as its thickness is negligible.

The peaked behavior of the inverse compressibility as a function of back-gate voltage or density may be used to map out the density fluctuations in the graphene sheet. At different locations across the sample the local density is added to the externally induced density by the back-gate. Therefore, different back-gate voltages are required in order to zero out the density at each location and reach the Dirac point. In practice this means that the entire inverse compressibility curve is shifted along the back-gate voltage axis as one moves from one point across the sample to another. Figure 2A shows a measurement of the inverse compressibility along an arbitrarily chosen line across the sample. The Dirac point is located within the red colored band, which corresponds to reduced compressibility. As expected, it appears at different back-gate voltages for different locations. The black dotted line shows the dependence of the Dirac point with position obtained by fitting the kinetic term of the compressibility to each back-gate voltage line scan. The smallest length scale on which density variations are observed is roughly 150 nm. This scale is most likely limited by our spatial resolution, i.e. the size of the SET. The underlying density fluctuations are therefore bound to occur on even smaller length scales and with higher amplitudes. Direct evidence for that is given in subsequent paragraphs.

The chemical potential variations within the graphene are likely due to charged impurities above and below the layer. In the process of scanning the SET above the layer one picks up not only the spatial variations in the chemical potential but also the potential emanating directly from the charges above the layer along with their image charges in the layer. As these potentials are much larger than the chemical potential variations it would be desirable to find means to subtract them out and be left with only the chemical potential variations. This can be readily achieved by subtracting off the measured potential at zero average density from that at a high carrier density. As the carrier density in the graphene sheet increases it is expected to screen better. Hence, at large carrier density the potential landscape in the graphene should be nearly constant and the only potential seen by the SET is that due to static charges above the layer. Using this method one can extract the local carrier density from the measurements of the surface potential only. A formal description of this subtraction process, which validates this method, is given in the supplementary material. Figure 2B shows a comparison of the density variations as extracted from the inverse compressibility measurements (black dotted line, which is identical to the one in Fig. 2A) with those extracted from the surface potential (blue solid line). There is striking quantitative agreement between the two methods. Fig. 3A depicts a two-dimensional map of the density variations of the graphene sheet when the average carrier density is zero. It was obtained from surface potential measurements. The red regions correspond to electrons and the blue regions correspond to holes. These data constitute direct evidence for the puddle model [13]. A statistical analysis of the density fluctuations present in this two-dimensional map is shown in Fig. 3B. The histogram plots the number of patches in which the density lies within a certain density interval. From a fit with a Gaussian distribution, we extract from its standard deviation that the density fluctuations are on the order of $\Delta n_{2D,B=0T} = \pm 3.9 \cdot 10^{10} cm^{-2}$.

The fluctuations in density discussed thus far have been resolution limited by our technique. However, the intrinsic density fluctuations may be extracted by going into the quantum Hall regime. A large magnetic field applied perpendicular to the sample produces bands of localized and extended states that are manifested in universal transport properties [33]. Our previous studies on GaAs two-dimensional electron systems have shown that at sufficiently large magnetic fields the width in density of the band of localized states, also referred to as the incompressible band, becomes field and also filling factor independent. This width constitutes a direct measure of the density fluctuations in the sample [34]. Fig. 4 shows a color rendering of the $\partial\mu/\partial n$ measured on our graphene flake as a function of the magnetic field and density. A single density scan at the fixed field of 11 T is included at the top. It is composed of a series of maxima at the integer fillings 2, 6, 10,…, corresponding to regions of low compressibility. These maxima can be fitted well by Gaussians of identical variance, which is in accordance with the filling independent width of the incompressible bands observed in conventional 2D systems. The variance may serve as a measure for the disorder amplitude. A best fit to the data is obtained assuming density fluctuations of approximately $\Delta n_{2D,B=11T} = \pm 2.3 \cdot 10^{11} cm^{-2}$. A similar analysis at lower magnetic fields shows that the variance still drops slightly with decreasing magnetic field and hence the value $\Delta n_{2D,B=11T}$ is a lower bound for the disorder strength as bands of localized states are not well separated yet. The disorder amplitude extracted from these measurements in the presence of a magnetic field is about a factor of six larger than the disorder estimate $\Delta n_{2D,B=0T}$ from the B = 0 T measurement in Fig. 3. This difference allows us to deduce the intrinsic disorder length scale, $l_{disorder}$. For the zero field estimate the density fluctuations are averaged over an area determined by the tip size with a characteristic dimension of approximately 150 nm (see Fig. 2B). The ratio of these averaged density fluctuations and the intrinsic ones is simply the square root of the ratio of the disorder area to the averaged area. Therefore we end up with an upper-bound for the intrinsic disorder length scale of about 30 nm. We note as a curiosity that a similar number for the disorder length scale is also obtained from the Einstein relation between conductivity and compressibility, $\sigma = e^2 D \frac{\partial n}{\partial \mu}$, when plugging in the expression for the diffusion constant valid for conventional 2D systems: $D = v_F l / 2$. Here, $l$ is the mean free path. The conductivity near the neutrality point is approximately $\sigma \approx \frac{4e^2}{h}$ and $\frac{\partial n}{\partial \mu}$ is shown in figure 1. We conclude that the intrinsic disorder length scale in graphene is approximately 30 nm. At high carrier density, the high compressibility of graphene smoothes out the disorder landscape. As one approaches the neutrality point however, screening becomes poor and the intrinsic disorder length turns relevant.

We acknowledge helpful discussions on the preparation of graphene flakes with K. Novoselov and A. Geim. We also would like to acknowledge fruitful discussions with F. von Oppen.

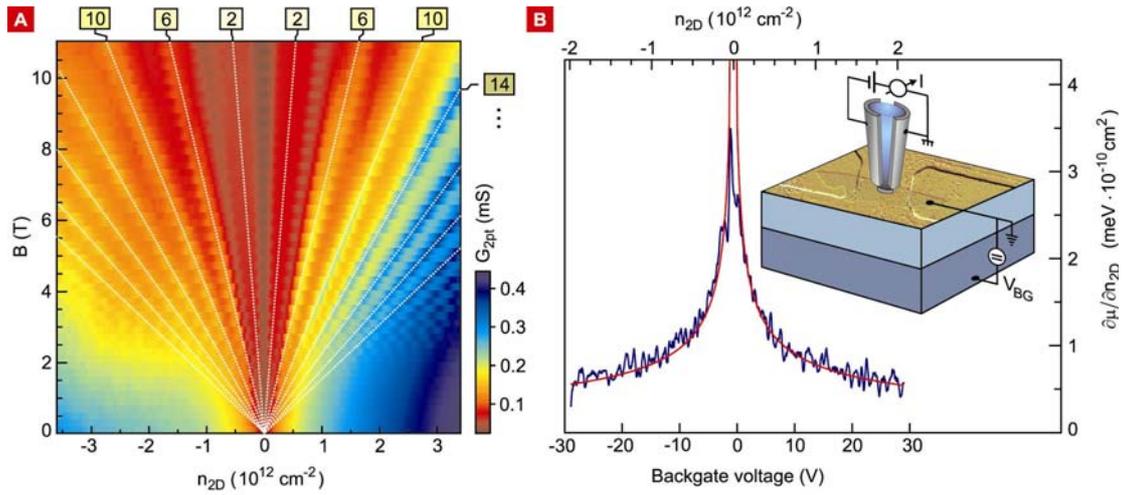

Fig. 1: A. Color rendering of the two-terminal conductance $G_{2pt}$ of a graphene monolayer in the density versus magnetic field plane. The conductance maxima follow slopes corresponding to integer fillings 2, 6, 10, …. , an unequivocal signature for a single monolayer. B. The inverse compressibility measured at an arbitrary, fixed location on the graphene sample as a function of the back-gate voltage or carrier density (blue line). The red line is a best fit to the data using an effective Fermi velocity as a single fit parameter in the kinetic energy contribution predicted from the graphene band-structure. The inset depicts the experimental arrangement consisting of an aluminum based SET evaporated on a glass fiber tip of a scanning probe microscope. The graphene monolayer is contacted with two leads on top of an oxidized and heavily doped Si wafer. A back-gate voltage induces charge carriers and modulates the density for the measurement of the local compressibility.

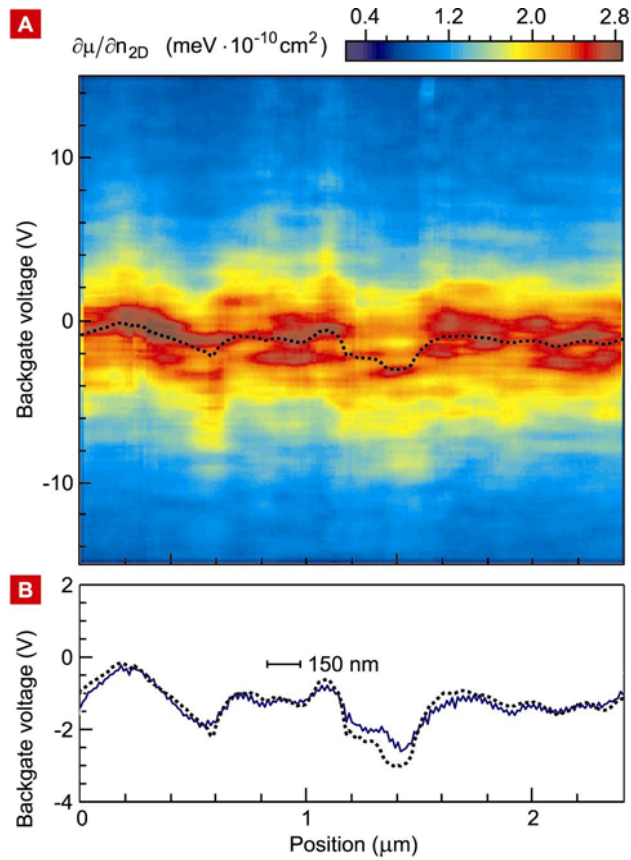

Figure 2: A. Color rendering of the inverse compressibility measured along a line across the sample as a function of the back-gate voltage. The dotted black line marks the location of the Dirac point obtained from a fit of the kinetic energy term to each line scan. B. Comparison of the spatial variation of the Dirac point extracted using two methods: the inverse compressibility (black dotted line, identical to the dotted line in panel A) and through subtracting surface potential scans at high and zero average carrier density (blue line). The bar marks the smallest length scale (approximately 150 nm) on which density variations are observed.

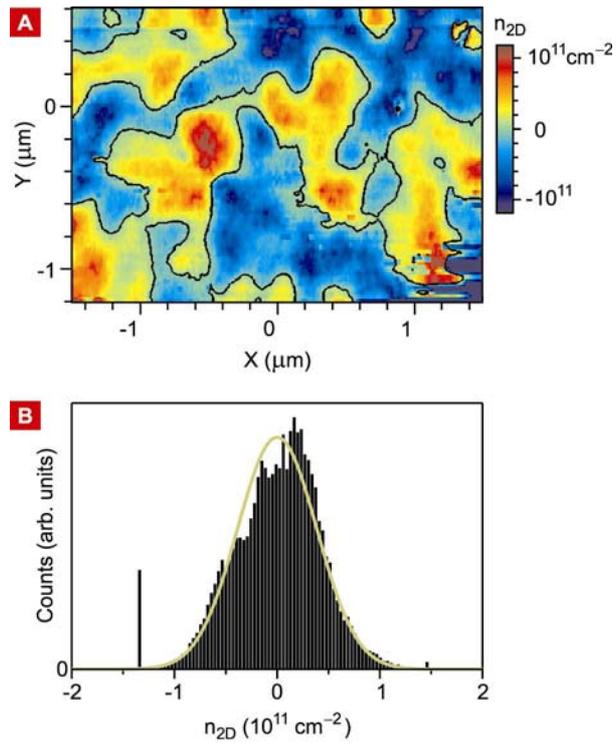

Fig. 3: A. Color map of the spatial density variations in the graphene flake extracted from surface potential measurements at high density and when the average carrier density is zero. Blue regions correspond to holes and red regions to electrons. The black contour marks the zero density contour. B. Histogram of the density distribution in A.

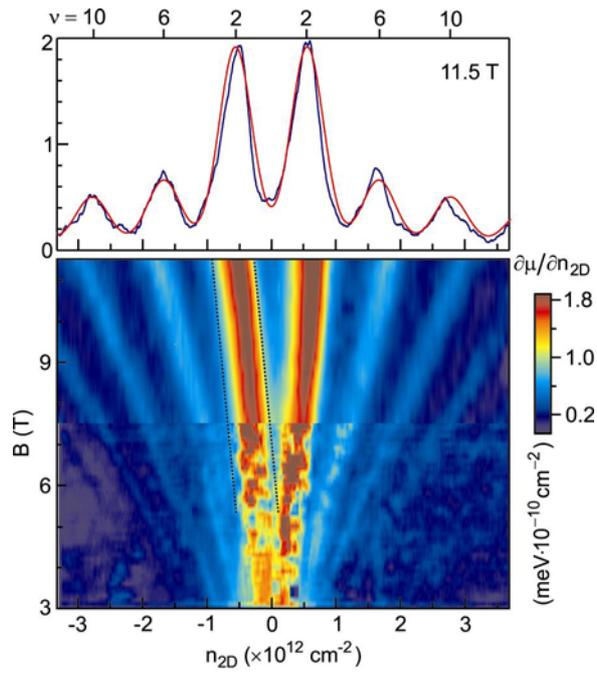

Fig. 4: A. Color rendering of the inverse compressibility as a function of density and magnetic field. B. A single line scan from plot A of the measured inverse compressibility (black line) at a magnetic field of 11 T. The red curve is a fit to the data composed of Gaussians with equal variance for each maximum. This variance measures the width of the incompressible regions (maxima) centered around integer fillings and provides an estimate for the intrinsic amplitude of the density fluctuations.

**Method Section**

*A. Determination of the local carrier density from the surface potential:*

Here we derive that a subtraction of the surface potential measured at high carrier density and at zero average carrier density yields the local density profile in the graphene flake. The most general expression for the surface potential detected by the SET reads

$$e\Phi_{total}(n,r) = e\Phi(r) + \mu(n,r), \tag{1}$$

where $\Phi(r)$ is the electrostatic potential due to local doping/adsorbates and $\mu(n,r)$ is the local chemical potential of the graphene sheet. The influence of disorder emanating from below and above the graphene flake on the local chemical potential is included in $\mu(n,r)$. We assume that $\mu$ only depends on the local density $n(r) = n_0 + \delta n(r)$. Here $n_0 = C_g V_{BG}$ is the average density in graphene which can be tuned with a back-gate and $\delta n$ is the disorder induced spatial density fluctuation. Close to the Dirac point where the average density $n_0 = 0$ equation (1) reduces to:

$$e\Phi_{total}(0,r) = e\Phi(r) + \mu(\delta n(r)) \tag{2}$$

At large voltages on the back-gate the average density will become larger than the density fluctuations $n_0 \gg \delta n$ and the surface potential may be written as:

$$e\Phi_{total}(n_0,r) \approx e\Phi(r) + \mu(n_0) + \frac{\partial \mu}{\partial n}(n_0) \delta n(r) \tag{3}$$

When we substract the average value of the surface potential measured at high electron density (positive back-gate voltage, Eq. (3)) and equal hole density (negative back-gate voltage) from the surface potential recorder close to the Dirac point (Eq. (2)), we derive:

$$e\Phi_{total}(0,r) - \frac{e\Phi_{total}(-n_0,r) + \Phi_{total}(n_0,r)}{2} = \mu(\delta n(r)) - \frac{\partial \mu(n_0)}{\partial n} \delta n(r)$$

$$\approx \left( \frac{\partial \mu(\delta n(r))}{\partial n} - \frac{\partial \mu(n_0)}{\partial n} \right) \cdot \delta n(r)$$

$$\approx \frac{\partial \mu(\delta n(r))}{\partial n} \cdot \delta n(r) \tag{4}$$

The density variations in the graphene sheet can then be obtained from:

$$\delta n(r) \approx \frac{e\Phi_{total}(0,r) - \dfrac{e\Phi_{total}(-n_0,r) + e\Phi_{total}(n_0,r)}{2}}{\dfrac{\partial \mu(\delta n(r))}{\partial n}} \qquad (5)$$

**B. Direct pick-up:**

The small size of the graphene flake and the finite distance between the single electron transistor (SET) and the graphene flake result in a directly measured potential contribution from the highly doped Si-back gate. The inset to Fig. 5 illustrates that some electrical field lines originating from the back-gate reach the SET. The parasitic charge associated with these fringing fields is also detected by the SET.

The detected amount of charge depends both on the distance of the SET to the graphene edge as well as on the vertical distance between tip and flake. Fig. 5 shows the dependence of the direct pickup measured as a function of height above a gold electrode. The gold electrode serves as a reference since it entirely screens the back-gate potential underneath and only the fringing fields are detected. The distance from the edge is chosen to be equal to the one in the real measurement on top of the graphene flake.

As expected, once the SET comes close to the surface the direct pick-up of the fringing fields drops and approaches zero. For typical heights and distances to the sample edge we derive a contribution from the fringing fields of approximately 0.15 meV $10^{-10}$ cm$^2$.

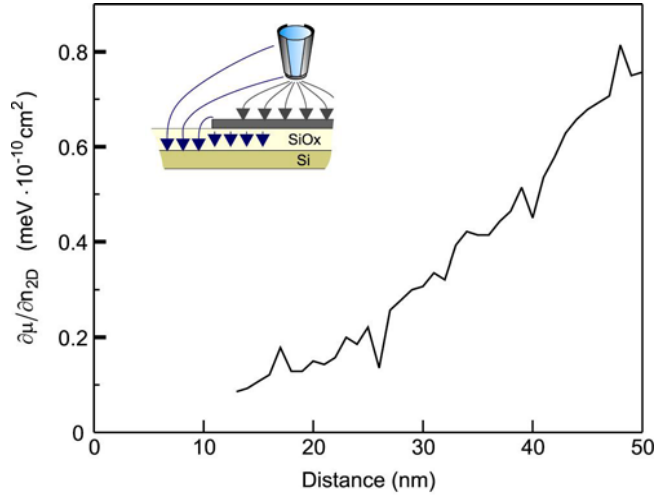

Fig. 5: The direct pick-up contribution to the measured inverse compressibility as a function of the distance between the tip and the sample. The tip is placed above a gold electrode at a fixed distance from the border of the electrode typical for the configuration in the experiment on a graphene flake. The inset schematically illustrates the origin of this direct pick-up.